\documentclass[graybox]{svmult}

\usepackage{helvet}         
\usepackage{courier}        
\usepackage{type1cm}        
\usepackage{makeidx}         
\usepackage{graphicx}        
\usepackage{multicol}        
\usepackage[bottom]{footmisc}

\usepackage{amsmath}
\usepackage{amsfonts}
\usepackage{amssymb}

\newcommand{\be}{\begin{equation}}
\newcommand{\ee}{\end{equation}}
\newcommand{\bea}{\begin{eqnarray}}
\newcommand{\eea}{\end{eqnarray}}

\newcommand{\eq}[1]{(\ref{#1})}

\def\a{\alpha} \def\ad{\dot{\a}} \def\ua{{\underline \a}}

\def\b{\beta}  \def\bd{\dot{\b}} \def\ub{{\underline \b}}

\def\d{\delta} 

\def\e{\epsilon}

\def\k{\kappa}

\def\l{\lambda}

\def\r{\rho}

\def\O{\Omega}

\def\cD{{\cal D}}

\def\cA{{\cal A}}

\def\cS{{\cal S}}

\def\cV{{\cal V}}

\def\cA{{\cal A}}

\def\yb{{\bar y}}
\def\zb{{\bar z}}

\def\tB{\widetilde{B}}
\def\tA{\widetilde{A}}

\def\bC{\boldsymbol{C}}

\def\bB{\boldsymbol{B}}
\def\bY{\boldsymbol{Y}}
\def\bZ{\boldsymbol{Z}}
\def\bX{\boldsymbol{X}}

\def\bcA{\boldsymbol{\cal A}}
\def\bcFS{\boldsymbol{\cal FS}}

\def\bcE{\boldsymbol{\cal E}}

\def\ovC{\overline{C}}

\def\mso{\mathfrak{so}}

\def\Real{{\mathbb R}}
\def\Comp{{\mathbb C}}

\def\ket#1{|#1\rangle}
\def\bra#1{\langle#1|}

\title*{Higher Spins, Holography and Exotic Matter}

\titlerunning{Higher Spins, Holography and Exotic Matter}

\author{Carlo Iazeolla and Per Sundell}

\institute{Carlo Iazeolla \at Department of Engineering Sciences, Universit\`a degli Studi Guglielmo Marconi, \\ Via Plinio 44, 00193, Roma, Italy \&
\\ Sezione INFN Roma Tor Vergata \\ Via della Ricerca Scientifica 1, 00133, Roma, Italy \\ Member of INDAM-GNFM \\
\email{c.iazeolla@gmail.com}
\and Per Sundell \at Instituto de Ciencias Exactas y Naturales, Universidad Arturo Prat, \\ Playa Brava 3265, 1111346 Iquique, Chile \&
\\
Facultad de Ciencias, Universidad Arturo Prat,\\ Avenida Arturo Prat Chacón 2120, 1110939 Iquique, Chile \\ 
\email{per.anders.sundell@gmail.com}}

\begin{document}

\maketitle

\abstract{We review a new perspective on higher-spin holography, whereby Vasiliev's 4D higher-spin gravity emerges together with a 3D counterpart, consisting of coloured conformal matter fields coupled to topological conformal higher-spin and colour gauge fields, as two distinct reductions, characterised by dual structure groups, of a common parent model. The latter is given by a non-commutative AKSZ sigma model consisting on-shell of a flat superconnection valued in a fractional-spin extension of Fradkin-Vasiliev's higher-spin algebra. In particular, we highlight an intermediate 4D reduction, referred to fractional-spin gravity, consisting of exotic matter, in the form of coloured singletons, coupled to higher-spin and colour gauge fields.}

\section{Introduction}
\label{sec:1}

Higher-spin gravities (HSGs) extend ordinary gravity by embedding the graviton into an infinite-dimensional multiplet containing gauge fields of all spins. Exotic as they may appear, HSGs are naturally connected to many branches of theoretical and mathematical physics. From a string theory vantage point, HSGs are expected to be fundamental, as higher-spin (HS) symmetries have been shown to emerge in suitable tensionless limits. In fact, the highly constraining power of their infinitely many gauge symmetries suggests their being renormalizable, or possibly even finite, models of quantum gravity (see, e.g., \cite{snowmass} and references therein).
Moreover, within holography, HSGs play a key role as candidate bulk duals of basic conformal field theories (CFTs) with $1/N$-expansions around free fields \cite{Sundborg:2000wp, Sezgin:2002rt,KP,LeighPetkou}.

At the same time, the distinctive features of HSG challenge the conventional QFT framework. For instance, as a consequence of the HS gauge symmetry, spacetime non-locality is expected to be part of a full interacting theory. Moreover, the usual concepts of Riemannian geometry lose any invariant meaning, prompting the development of an appropriate HS generalization of geometry in order to capture the physical content of HSG. It is hard to tackle such issues within the usual field theoretic setup.  In fact, already the infinite-dimensional symmetry coupled with its homotopy Lie algebra structure forced the introduction of a more general framework in order to write fully non-linear equations. Such framework blends the formalism of $Q$-manifolds and non-commutative geometry to encode the complicated $L_\infty$ structure of HSG into a compact set of first-order constraints known as the Vasiliev equations \cite{MV,properties,more,review99,solvay,SlavaZhenya,formal}. The mathematical structures involved unlock new tools for handling non-localities, either by suitable relaxations of locality \cite{spinloc1,spinloc2,spinloc3,spinloc4,spinloc5,spinloc6,spinloc7} or by sidestepping the problem with the construction of geometric observables defined on Vasiliev's non-commutative extension of the spacetime base manifold \cite{Sezgin:2011hq,NiccoPer1,NiccoPer2,FSG1,FSG2}.  

It is therefore very interesting to reconsider holography within Vasiliev's HSG framework. A key feature thereof is that classical solutions describe non-commutative geometries, i.e.,  fundamental fields belong to differential graded associative algebras (DGAs) consisting of operators acting in modules with projections to representations dual to boundary conditions; for example, asymptotic anti-de Sitter regions are dual to holomorphic extensions of metaplectic group algebras, which thus arise already at linearized level. Such spaces of classical solutions, equipped with additional Poisson structures introduced via a sigma model of Frobenius-Chern-Simons (FCS) type \cite{FCS1,FCS2}, may then be quantized by exploiting the AKSZ formalism \cite{AKSZ}. A suitable extension of the HS algebra to a \emph{fractional-spin algebra} turns the FCS system into a parent model encompassing both 4D HSG and a specific 3D CFT as consistent reductions --- a direct relation between candidate holographic dual pairs which lies at the heart of an approach to HS holography proposed in \cite{FSG1,FSG2}.

What follows is a review of the results obtained in \cite{FSG1,FSG2}. Section 2 is a lightning review of the Vasiliev system and its FCS extension. Section 3 outlines the main ideas and results of the HS/CFT correspondence. In Section 4, we describe the parent model, and show how it contains 4D HSG as well as a 3D candidate dual, consisting of coloured conformal fields coupled to topological higher-spin and colour gauge fields (CCHSG), as consistent reductions. Finally, we review an intermediate consistent reduction describing a novel 4D system, which we refer to as \emph{fractional-spin gravity} (FSG), describing the coupling of HSG to exotic matter in the form of coloured singletons, after which we provide a short outlook.

\section{Higher-spin gravity and Frobenius-Chern-Simons superconnections}
\label{sec:2}

In this Section we recall the basics of the 4D bosonic Vasiliev system and of its Frobenius-Chern-Simons extension, as a starting point for our construction.

\subsection{Higher-spin algebras}\label{sec:Vas}

The basic algebraic structures of holography, i.e, ensuing groups and representations organising 3D conformal as well as 4D gauge theories with HS symmetries, are hard-wired into Vasiliev's HSGs, whose Lorentz-covariant classical perturbative expansions around $AdS_4$ backgrounds are governed by associative gauge algebras, defined originally via oscillator realisations \cite{FV,Konstein:1989ij}, and subsequently via quotienting enveloping algebras by Joseph ideals \cite{Eastwood:2002su,MValg,fibre}. The gauge algebra of pure HSG is the quotient
\be
\boldsymbol{\mathcal{HS}}=\frac{\boldsymbol{{\rm Env}}(\mathfrak{so}(2,3))}{\boldsymbol{\mathcal{I}}({\cS})}\ ,
\ee
of the unital enveloping algebra of the Lie algebra $\mathfrak{so}(2,3)\cong \mathfrak{sp}(4,\Real)$, i.e., the space of polynomials in $\mathfrak{so}(2,3)$-generators extended by a unit, and the annihilating ideal $\boldsymbol{\mathcal{I}}(\mathcal{S})$ of the singleton $\cS$. The latter is an $\boldsymbol{\mathcal{HS}}$-module  that can be completed into representations of the metaplectic group $Mp(4;\Real)$ and its complexification, corresponding to different boundary conditions on 3D conformal scalars and spinors. 

$\boldsymbol{\mathcal{HS}}$ can be realized as the Weyl algebra $A_2 \cong (\Comp[Y^{\ua}],\star)$ represented by symbols given by even polynomials in an $Sp(4,\mathbb{R})$ quartet $Y^{\ua}$ --- splitting under $SL(2,\mathbb{C})$ into complex two-component spinors, viz., $Y^{\ua}=(y^\a,\yb^{\ad})$ --- composed using the Moyal star product $\star$; in particular, one has the Heisenberg algebra relations 
\be  [Y^{\ua},Y^{\ub}]_\star=Y^{\ua}\star Y^{\ub}-Y^{\ub}\star Y^{\ua}=2i\left(\begin{array}{cc} \e^{\a\b} & 0 \\ 0 & \e^{\ad\bd} \end{array} \right)\ .\ee
However, the Lie algebra induced from $\boldsymbol{\mathcal{HS}}$ admits no finite-dimensional subalgebras beyond $\mso(2,3)$. Hence, the HS gauge connection must be expanded over generators of arbitrarily high spin \cite{solvay}. In other words, the perturbative treatment of HS gauge symmetries forces extensions of $A_2$ into non-polynomial completions that remain associative.

Imposing asymptotic boundary conditions on $AdS_4$ leads to classical perturbative expansions contained in the specific non-polynomial extension \cite{fibre,COMST,meta,corfu21,FSG1}
\be
\boldsymbol{\cal A}=\boldsymbol{\cal K}\star \left(\boldsymbol{\cal G} \oplus \boldsymbol{\cal G}^{(\infty)}\right)\ ,
\ee
with indecomposable product structure $\boldsymbol{\cal G}\star \boldsymbol{\cal G} = \boldsymbol{\cal G}$, $\boldsymbol{\cal G} \star \boldsymbol{\cal G}^{(\infty)}=\boldsymbol{\cal G}^{(\infty)}$ and $\boldsymbol{\cal G}^{(\infty)} \star \boldsymbol{\cal G}^{(\infty)}=\boldsymbol{\cal G}^{(\infty)}$, where 
\begin{itemize}

\item $\boldsymbol{\cal G}$ is the holomorphic oscillator realization of the algebra of the complex inhomogeneous metaplectic group $MpH(4;\Comp)$ \cite{meta}, viewed as a branched double cover of $SpH(4;\Comp)$, containing gauge functions from $Mp(4;\Real)$ and proper holomorphic elements in $Mp(4;\Comp)\setminus Mp(4;\Real)$ providing complexified Bogolyubov transformations that connect different boundary conditions of 3D conformal and 4D HS fields, e.g., the inner Kleinians $(\k_y,\bar\k_{\yb})$ which are mutually commuting unimodular operators obeying 
\be
\{\k_y,y_\a\}_\star = 0 = \{\bar \k_{\yb},\yb_{\ad}\}_\star \ ;\label{ky}
\ee
\item $\boldsymbol{\cal G}^{(\infty)}$ is a two-sided $\boldsymbol{\cal G}$-module arising at the asymptotic boundary of $\boldsymbol{\cal G}$ \cite{meta}, representing ramification points of $SpH(4;\Comp)$ \cite{meta} corresponding to projectors onto representations inside $\cS$ (e.g., the unitarizable compact weight spaces of bounded conformal fields on $S^1 \times S^2$, squaring to massless particles in $AdS_4$ \cite{FF}, and non-compact counterparts arising on $CMink_3$, whose squares contain bulk-to-boundary propagators \cite{Corfu19,FSG1}); and
\item $\boldsymbol{\cal K}$ is a discrete group algebra generated by outer Klein operators $(k,\bar k)$ adding sectors with twisted boundary conditions to the underlying first-quantised system \cite{meta}, viz., $\{k,y_\a\}_\star = 0 = \{\bar \k,\yb_{\ad}\}_\star$.

\end{itemize}

\subsection{Vasiliev's equations}

Vasiliev's theory \cite{Vasiliev90,properties,more,review99,solvay,SlavaZhenya} combines two basic ingredients:
\begin{itemize}
\item The \emph{unfolding} \cite{MV} (see also \cite{review99,solvay}) of partial differential equations (PDE) whereby PDEs are converted into universally Cartan integrable systems (CIS) of constraints on the exterior derivatives of sets of differential forms\footnote{CIS arise naturally as equations of motion of AKSZ sigma models \cite{AKSZ,Barnich:2009jy,formal,(20)} by converting forms into maps from one set of $Q$-manifolds, referred to as the sources, into another, referred to as the targets.
Letting the DGA of functions on the target be generated from graded coordinates $X^\alpha$, and expanding the target $Q$-structure as $\vec Q=Q^\alpha(X)\vec \partial_\alpha$, the equations of motion read
$dX^\alpha=Q^\alpha(X)$, with $X^\alpha$ treated as forms on the source, forming a CIS by virtue of $\vec Q^2=0$.}, viewed as fundamental fields;, 
\item \emph{Non-commutative geometries} as described by DGAs with symbols given by differential forms on first-quantised differential Poisson manifolds. 
\end{itemize}
Vasiliev's fundamental fields are horizontal forms on fibered non-commutative manifolds 
$ \bY\rightarrow \bC \xrightarrow{p} \bB$
referred to, respectively, as \emph{master fields} and \emph{correspondence spaces}.
The bases $\bB$ provide the set of sources of an AKSZ sigma model, with classical moduli spaces obtained by taking $\bB\cong \bX_4\times \bZ$,
where $\bX_4$ is a spacetime manifold and $\bZ$ a non-commutative differential Poisson manifold supporting closed and central elements; correspondingly, the horizontal de Rham differential $d=d_{X} + d_{Z}$.
Manifestly Lorentz-covariant models arise by taking $\bZ$ to be a compact version of $\bY$, i.e., the projection of $\boldsymbol{\Omega}(\bZ)$ into top-form degree is represented by integrable symbols.
To represent the algebra $\boldsymbol{\mathcal{A}}\otimes \boldsymbol{\Omega}(\bZ)$ using symbols, we introduce global canonical coordinates $Z^{\ua}=(z^\a,-\zb^{\ad})$ on $\bZ$, obeying
\be [Z^{\ua},Z^{\ub}]_\star \ = \ -2i \left(\begin{array}{cc} \e^{\a\b} & 0 \\ 0 & \e^{\ad\bd} \end{array} \right) \ ,  \qquad [Y^{\ua},Z^{\ub}]_\star \ = \  0 \ ;\ee
choose a normal ordering with respect to the combinations $Y\pm Z$, in which symbols from $L^1(\bY\times \bZ)\cap L^2(\bY\times \bZ)$ are composed using 
\be f(Y,Z)\star g(Y,Z) = \int\frac{d^4U d^4V}{(2\pi)^4} \,e^{iVU}\,f(Y+U,Z+U)g(Y+V,Z-V) \ ,\ee
where $VU:=V^{\ua}U_{\ua}$; and extend to a holomorphic metaplectic group algebra containing inner Klein operators $(\k_z, \bar \k_{\zb})$ obeying analogues of \eq{ky}.
Finally, the intertwining rules for outer Kleinians are extended to $\boldsymbol{\mathcal{A}}\otimes \boldsymbol{\Omega}(\bZ)$ by taking $\{k,z_\a\}_\star = 0 = \{\bar \k,\zb_{\ad}\}_\star$.

Introducing the projected\footnote{The projection removes half-integer spins \cite{review99}; in Weyl order, the system can be further restricted to $\left(\Pi_{\rm bos} \star\boldsymbol{\Omega}(\bB)\star \Pi_{\rm bos}\right) \otimes \left(\Pi_{\rm bos} \star\bcA\star \Pi_{\rm bos}\right)$. } DGA 
\be \bcE_{\rm hor}(\bC):=\Pi_{\rm bos} \star\boldsymbol{\Omega}_{\rm hor}(\bC)\star \Pi_{\rm bos}\ ,\ee
with $\Pi_{\rm bos}:=\frac{1+k\bar k}{2}$, the Vasilev system consists of:
\begin{itemize}
\item a dynamical one-form $A\in\bcE_{1\,\rm hor}(\bC)$ whose projection to $H(d_Z)$ is represented by $\check A :=A|_{Z=0}\in\Omega_1(\bX_4)\otimes\bcA$ being a \emph{connection one-form} in the \emph{adjoint} representation of $\boldsymbol{\mathcal HS}$ \cite{review99} containing of gauge fields of all spins;  
\item a dynamical zero-form $B\in\bcE_{0\,{\rm hor}}(\bC)=\bcE_{0}(\bC)$ whose projection to $H(d_Z)$ is represented by $\check B :=B|_{Z=0}\in\Omega_0(\bX_4)\otimes\bcA$ containing a \emph{Weyl zero-form} $\Phi:=(B\star k)|_{k=\bar k=0}$ in the \emph{twisted-adjoint} representation of $\boldsymbol{\mathcal HS}$ \cite{review99}, containing a scalar, Faraday tensor, Weyl tensor and HS generalizations thereof; 
\item the cohomologically non-trivial, closed and central two-form
\be I_\Comp :=  e^{i\theta_0}\,j_z \star \k_y \star k - {\rm h.c.} \ ,  \ee
where the manifestly $SL(2,\mathbb{C})$-invariant holomorphic two-form
\be j_z:= -\frac{i}4\, dz^\alpha\wedge dz_\alpha\, \kappa_z\ ,\ee
and the phase $\theta_0$ breaks parity, except for $\theta_0=0$ and $\pi/2$, for which the scalar is even and odd, respectively, under parity \cite{Sezgin:2003pt}.
\end{itemize}
In terms of these forms, the Vasiliev equations take the remarkably compact form
\bea   
dA + A\star A + B\star I_{\Comp} &=& 0 \label{Vas1}\\
dB+A\star B-B\star A &=& 0 \ ,\label{Vas2}
\eea 
whose Cartan integrability, i.e., compatibility with $d^2=0$, implies the gauge symmetries
\be \d A \ = \ d\e + [A,\e]_\star \ , \qquad \d B = [B,\e]_\star \ . \ee
Viewing $\bC$ as a non-commutative background, the system describes how the closed and central element $I_\Comp$ combines with $B$ into a source $B\star I_\Comp$ for the curvature of $A$, i.e., a non-linear, field-dependent deformation of the differential Poisson structure of $\bC$, dual to deformation of the DGA structure of $\bcE_{\rm hor}(\bC)$.
Moreover, the $SL(2,\Comp)$-invariance of $I_\Comp$ ensures that a canonical Lorentz connection on $\bX_4$ can be embedded into $A$ hence $\check A$ (see \cite{FSG2}, based on previous analysis in \cite{properties,review99,analysis}).

Solving the components of \eq{Vas1}-\eq{Vas2} in the co-kernel of $\imath_{\vec\partial^{(Z)}_{\ua}}$ subject to initial conditions at $Z=0$, the remaining constraints in the kernel of $\imath_{\vec\partial^{(Z)}_{\ua}}$, hold for all $Z$ if they hold at $Z=0$, generating a CIS for $\check A$ and $\check B$ on $\bX_4$ based on a strong homotopy $L_{\infty}$  algebra, viz.,
\bea    
d_X\check A &= & - \check A\star \check A + {\cal V}^{(1)}(\check A,\check A, \check B) + {\cal V}^{(2)}(\check A,\check A, \check B, \check B) + ...\label{VasV}\\
d_X\check B &=& - [\check A, \check B]_\star + {\cal U}^{(2)}(\check A,\check B, \check B) + ... \label{VasU}\eea
where ${\cal V}^{(n)}$, $n=1,2,\dots$, and ${\cal U}^{(n)}$, $n=2,\dots$, are structure maps comprising interactions of order $n$ in $\check B$, deforming the gauge symmetries of $\check A$ and $\check B$ in a compatible fashion. 
In this sense, $\bcE_{\rm hor}(\bC)$, viewed as a DGA with internal differential $d_Z$ fibered over $\bX_4$, constitutes a cohomological resolution of the minimal $L_\infty$ model \eq{VasV}-\eq{VasU} defined on $H(\bcE_{\rm hor}(\bC),d_Z)$, with $L_\infty$ structure transferred to the $d_Z$-cohomology via homological perturbation theory \cite{ht,formal,LiZeng}.

\subsection{Frobenius-Chern-Simons extension}
\label{sec:FCS}

As we have recalled, the source of the Vasiliev system, which deforms the symplectic structure on $\bZ$, is built via the non-dynamical two-form $I_\Comp$. It is actually possible to generalise the Vasiliev system to include a dynamical two-form $\tB$, in such a way that Eqs. \eq{Vas1}-\eq{Vas2} are recovered as a particular reduction of the complete system, referred to as Frobenius-Chern-Simons (FCS) gauge theory, triggered by the specific vacuum value $I_\Comp = \langle \tB \rangle$  \cite{FCS1,FCS2}. The main benefit of such an extension is that it is possible to write an action of Chern-Simons/AKSZ sigma-model type \cite{AKSZ} for this generalised system via the addition of bulk Lagrange multipliers. Furthermore, while the FCS theory propagate the same local degrees of freedom, it has an enlarged gauge symmetry, reducing greatly the number of HS gauge invariant observables.

The whole system lives on a correspondence space $\bY\rightarrow \bC_{\rm ext} \xrightarrow{p} \bB_{\rm ext} $ extended with one extra spacetime dimension $\bB_{\rm ext} \cong \bX_5\times \bZ$, such that the base of the correspondence space of the Vasiliev system is the boundary of this extended space, $\bX_4  = \partial \bX_5 $ . Moreover, it is assumed that $\bZ$ is compact, $\bZ\cong S^2_{\Comp}\times \bar S^2_\Comp$, in such a way that the algebra of differential forms on $\bZ$ has a well-defined trace operation projecting onto top forms (used in defining action and Chern classes)\footnote{For this reason, the FCS/AKSZ model is equivalent to the Vasiliev system locally but not globally, i.e. not on the entire noncommutative geometry including boundary conditions on $\bZ$.}. Cartan-integrability of the extended system requires a new one-form connection $\tA$ to be added to $A$. Moreover, $B$ and $\tB$ are assumed to transform in the bifundamental representation. 

Within this extended FCS model, the Vasiliev system \eq{Vas1}-\eq{Vas2} is retrieved, upon imposing a specific gauge choice and AKSZ-inspired boundary conditions on the Lagrange multipliers, in the form of \emph{boundary equations} on $\partial \bB_{\rm ext}$ \cite{FCS1,FCS2,FSG1}. The latter take the form of a flatness condition for a superconnection $X$ in which the four master-fields $A$, $\tA$, $B$, $\tB$ are assembled. $X$ is in fact valued in a $3$-graded Frobenius algebra ${\rm mat}_{1|1}$, generated by elements $m_{ij}$ such that
\be {\rm mat}_{1|1}=\bigoplus_{i,j=1,2} \mathbb{C}\otimes m_{ij} \ , \quad {\rm deg}(m_{ij})=j-i 	\ ;\ee
$X\in {\rm mat}_{1|1}\otimes\bcE_{\rm hor}(\bC_{\rm ext})$ and of total degree one, in such a way that 
\be X=m_{11}\star A+m_{22} \star\tA+m_{12}\star B+m_{21}\star \tB \equiv \left[\begin{array}{cc} A & B\\  \tB  &\tA \end{array}\right]\ .\label{mat11} \ee
Thus, the flatness condition on $X$,
\be dX+X\star X = 0 \ , \label{Xflat}\ee
contains, sector by sector, the equations (using Koszul sign convention for tensoring ${\rm mat}_{1|1}$ with differential forms)
\bea  dA+A\star A+B\star \tB &= &0\ ,\qquad d\tA+\tA\star \tA+\tB\star B= 0\ ,  \label{Xflat1}\\
 dB+A\star B-B\star \tA&=& 0\ ,\qquad d\tB+\tilde{A}\star \tB-\tB \star {A}= 0 \ , \label{Xflat2}\eea
which for $A=\tA$ and $\tB=I_\Comp$ indeed reproduce the Vasiliev equations \eq{Vas1}-\eq{Vas2}. Reality conditions can be imposed on $X$ such that $A^\dagger = -A$, $\tA^\dagger = -\tA$, $B^\dagger=B$, $\tB^\dagger=-\tB$. Both the Frobenius algebra extension and the formulation  on a noncommutative base manifold with boundaries will be instrumental in the formulation of a parent model for holographically dual HSG theories. 

\section{Interlude: HS/CFT conjectures}
\label{sec:3}

Maldacena's holography conjectures, made in the context of large-$N$ expansions of coinciding branes, were predated by proposals \cite{Bergshoeff:1988jm} treating bilinear operators of 3D CFTs arising on single supermembranes on $AdS_4\times S^7$ as vertex operators for 4D HSGs.
These ideas were reanimated in \cite{Sezgin1998}, extrapolated in \cite{Sundborg:2000wp} to tensionless superstrings on $AdS_5\times S^5$, dressed by $1/N$-expansions in \cite{WittenJHS60,Sezgin:2002rt}, and subsequently reduced to IR fixed points in \cite{KP,LeighPetkou}.
These works suggest that 4D HSG is holographically dual to $1/N$-expansions of singlet sectors of 3D CFTs built from conformal scalars and spinors, whose HS currents are conserved to leading order, whereby the asymptotically free bulk HSG gauge fields provide sources for conserved HS currents built from bilinear or sesquilinear constructs of free CFT fields\footnote{Maldacena and Zhiboedov \cite{MZh} showed that a unitary, local, 3D CFT containing at least one conserved HS current must be free.}.

Starting from the HSG-side, tests were performed by solving the equations on $\bZ$, computing quadratic vertices on $\bX_4$ for Fronsdal fields, and extracting three-scalar \cite{Sezgin:2003pt} and more general \cite{GiombiYin,GiombiYin2} three-point functions using the on-shell GKPW prescription. However, this approach to holography in the HSG context is problematic: 

\begin{enumerate}
\item Obtaining the $Z$-dependence using generic contracting homotopies may yield space-time non-localities already at the level of cubic vertices, causing divergencies in boundary correlators\footnote{This prompted the development of the spin-local scheme, producing well-defined cubic vertices, holographically tested in \cite{Didenko:2017lsn}; see also \cite{Sezgin:2017jgm} for a related scheme. Alternatively, chiral HSGs (see e.g. \cite{Ponomarev:2016lrm,Sharapov:2022faa,Sharapov:2022awp,Didenko:2022qga}) have local vertices yielding three-point functions coinciding with those of non-chiral HSG \cite{Skvortsov:2018uru}.}  \cite{Boulanger:2015ova};
\item The existence of a Lagrangian density for HSG written solely in terms of Fronsdal fields on $\bX_4$ remains an open problem, which further complicates implementing the usual AdS/CFT prescriptions.
\end{enumerate}
Starting instead from first principles, as for any set of PDEs, Vasiliev's equations become predictable only when supplemented with boundary conditions on $\bC$, i.e., a choice of representation for  $\boldsymbol{\mathcal E}_{\rm hor}(\bC)$.
The resulting classical solution spaces can be probed using gauge-invariant functionals given by integrals over $\bZ$ that are sensitive to the choice of boundary conditions but not to the specifics of the contracting homotopy (as long as these act faithfully on $\boldsymbol{\mathcal E}_{\rm hor}(\bC)$).
It has been shown \cite{NiccoPer1,NiccoPer2,SlavaZhenya} that boundary correlators can be extracted from specific invariants of the Vasiliev system.
In particular, the second Chern class
\be \int_{\bZ} {\rm Tr}_{{\bcA}} (dA+A\star A)^{\star 2} \approx {\rm STr_{\bcA}} (\Phi\star\kappa_y\star \Phi\star \kappa_y)  \ .\ee
For instance, letting $\Phi_i$ be the master-field corresponding to the bulk-to-boundary propagator with one leg on the boundary point $x_i$, this quantity reproduces the two-point function of boundary currents in the free CFT, 
\be  {\rm STr_{\bcA}} (\Phi_1\star\kappa_y\star \Phi_2\star \kappa_y) \ = \ \langle {\cal J}_1{\cal J}_2\rangle_{free\, CFT} \ .\ee
In \cite{COMST}, it has been proposed that imposing asymptotic boundary conditions in $\Phi_i$ yields higher-order corrections to the second Chern class containing $n$-point functions probing Vasiliev's theory. 
These results indicate that viewing the full Vasiliev system as a non-commutative gauge theory on $\bC$ --- for which the AKSZ formalism provides a natural framework --- may bypass step 1 and bring some revision of the HS/CFT conjecture.

Another reason to expect a revision is that HSG admits a number of solutions --- e.g. HS black-hole- and black-brane-like solutions \cite{Didenko:2009td,2011,biaxial,2017,brane1,brane2}, vacua with flat-but-nontrivial $Z$-space connections \cite{Iazeolla:2007wt}, and $BTZ_4$-like vacua encoding degenerate asymptotic conformal geometries \cite{BTZ} --- that introduce moduli that challenge the ordinary formulation of the dual 3D CFTs using Lagrangian densities on conformal spacetimes (e.g., critical and non-critical vector models \cite{MZh2} and vector models with Chern-Simons interactions \cite{vecCS}).

Indeed, viewing the HSGs as AKSZ sigma models suggests similar reformulations of the 3D CFTs.
In this respect, an indication comes from the results of \cite{Misha}, where Vasiliev found that the restriction of 4D unfolded linearised HS field equations on a 3D leaf of the Poincar\'e patch of $AdS_4$ produces unfolded equations for conformal topological HSG fields coupled to conformal currents, together with conservation conditions for the primary currents. 
As spacetime is largely ``pure gauge'' in the unfolded formulation, it was proposed that holographically dual systems admit isomorphic unfolded formulations. Thus, the observed reduction led to the conjecture that the actual holographic dual of 4D HSG is 3D conformal HSG (CHSG) \cite{Pope:1989vj,F&L,Segal,Tseytlin:2002gz} coupled to conformal matter (see also \cite{Dio+Vasilis} for a similar result, obtained by different means, on euclidean de Sitter background). The fact that the dual boundary theory is not free does not contradict the Maldacena-Zhiboedov theorem, since, being CHSG a gauge theory, it escapes at least one of the assumptions of unitarity, locality and/or conformal invariance. However, this ``bottom-up'' approach was intrinsically classic, on-shell and linear. Moreover, conformal currents appear this way as primary dynamical object, not directly as matter constructs.

As we shall see next, a generalization of the FCS/AKSZ approach gives a chance of framing the holographic duality in an independent, ``top-down'' approach, building both dual systems directly in correspondence space (i.e., both are non-linear theories). 
Indeed, a generalization of the algebra that the parent superconnection $X$ is valued in gives rise to a parent model for both 4D HSG and a specific 3D candidate dual theory: a CHSG coupled to coloured conformal matter fields and colour gauge fields, which we shall refer to as \emph{coloured conformal HSG} (CCHSG).

\section{Parent model and reductions}
\label{sec:4}

As we have seen in Section \ref{sec:FCS}, 4D HSG can be obtained as boundary equations of an FCS/AKSZ model. The particular reduction of the system leading to \eq{Vas1}-\eq{Vas2} is triggered by $\langle \tilde{B}\rangle = I_{\mathbb{C}}$, cohomologically non-trivial central two-form, with $Z$-dependent ``core'' factors $j_z$ and $\bar j_{\zb}$ reducing the structure group to their stabilizer $SL(2,\Comp)$.

It turns out, however, that the FCS model can accommodate different consistent reductions via an appropriate generalisation of the fibre algebra: in particular, we shall see that different theories --- including theories with different field content and living on spacetime manifolds of different dimensions, such as 4D HSG and 3D CCHSG --- get a common parent model corresponding to the boundary field equations of an AKSZ model. In this sense, \emph{solution spaces of different classical theories are interpreted as different spaces of boundary states of a common AKSZ parent theory}. Besides relating them directly, such an embedding into a parent model also provides an \emph{a priori} rationale for establishing holographic relations between invariant functionals of the two dual theories. 
This idea is at the core of the envisaged AKSZ-inspired approach described more in detail in \cite{FSG1}.

\subsection{Fractional-spin algebra extension}\label{sec:FS}

The relevant fibre algebra extension amounts to taking $X\in \bcE_0(\bB)\otimes{\rm mat}_{1|1}\otimes {\bcFS}$, where $\bcFS$ is a \emph{fractional-spin algebra}. 3D algebras of this type were introduced in \cite{frac1,frac2} in order to account for fractional-spin degrees of freedom, and generalized in \cite{FSG1} as algebras of endomorphisms of projected, first-quantized two-particle systems. A concrete realization that is relevant for modelling holographic correspondences involving higher spins on $AdS_4$ is the algebra
\be  \bcFS \cong {\rm {End}}(\cS \oplus \cV ) \ .\ee
of endomorphisms of the direct sum of two left-modules: a conformal singleton module $\cS$ and an internal, finite-dimensional Hermitian module $\cV = {\rm span}_{\Comp}\{|e_+^I\rangle, |e_-^{I'}\rangle\}_{I=1, ...N_.; I'=1,...,N_-}$. The singleton, which is an $Mp(4,\Real)$ module, decomposes into $SL(2,\Comp)$ submodules, as such consisting of various fractional-spin representations dictated by boundary conditions\footnote{For example, the compact and conformal scalar singletons \cite{FSG1}, $\cD(1/2,0)$ and ${\cal T}(-i/2,0)$ respectively, remain $SL(2,\Comp)$-irreps induced from singlets of $SO(3)$ and $SO(1,2)$, both with quadratic Lorentz Casimir given by $-3/4$, providing examples of fractional-spin representations.}.  $\cV$ is also equipped with an inner product $\left(| e^I_\epsilon \rangle, |e^{I'}_{\epsilon'}\rangle\right)_{\cV}:=\epsilon \delta_{\epsilon,\epsilon'}\delta^{II'}$, in general with split signature $(N_+,N_-)$, preserved by $U(N_+,N_-)$.

The inclusion of $\cV$ brings in Chan-Paton, colour-like degrees of freedom whose role will become clear later. In the specific consistent reductions of interest, depending on the form of the central element triggering the reduction, $\cV$ may or may not be allowed to break the modular symmetries generated by the Klein operators: in the first case we can always restrict to a positive definite space, i.e., a Hilbert space with norm-preserving endomorphism group $U(N)$, at the expense of breaking the modular element $\kappa_y$.

Thus, each block of the superconnection $X$ \eq{mat11} is now valued in $\bcFS$. We shall for this reason notationally distinguish its elements with boldfaced characters, 
\be \label{XFS}
\left.X\phantom{)}\right\downarrow_{{\rm mat}_{1|1}} = \left[\begin{array}{c|c}\mathbb{A}&\mathbb{B}\\\hline\widetilde{\mathbb{B}}&\widetilde{\mathbb{A}}\end{array}\right]\ ,
\ee
each of which decomposes under $\bcFS$ into a $2\times 2$ matrix as 
\be \label{FSdec}
\left. (\mathbb{A}, \ \widetilde{\mathbb{A}}, \ \mathbb{B},\ \widetilde{\mathbb{B}}) \ \right\downarrow_{\boldsymbol{{\cal FS}}} \ \sim \ \left[\begin{array}{c|c}\ket{\cS}\bra{\cS} &\ket{\cS}\bra{\cV}\\\hline\ket{\cV}\bra{\cS}\ &\ket{\cV}\bra{\cV}\end{array}\right]  \ ,\ee
meaning that the first element on the diagonal is expanded over symbols of operators realising endomorphisms of $\cS$ (the shorthand notation $\ket{\cS}\bra{\cS}$ denoting elements of ${\rm End}(\cS)$); the second diagonal element is expanded over symbols of operators realising ${\rm End}(\cV)$; while off-diagonal elements are expanded over elements that are intertwiners, transforming under the external symmetry algebra $\cA$  on one side and the internal $U(N_+,N_-)$ on the other side. 

Elements of all sectors admit a star-product realisation, so they can be represented by functions of $Y$ forming a star-product algebra. In particular, mixed bimodules elements can be obtained via star products of pure bimodules, in a unique way (up to normalizations), i.e., schematically, $\ket{\cS}\bra{\cV} = \ket{\cS}\bra{\cS} \star \ket{\cV}\bra{\cV}$ \cite{FSG2}. 

The interplay of the ${\rm mat}_{1|1}$ and the $\bcFS$ algebra is crucial to embed models with different dynamics via different consistent truncations of $X$. 
Indeed, as
\bea 
X=\left[\begin{array}{c|c}\mathbb{A}&\mathbb{B}\\\hline\widetilde{\mathbb{B}}&\widetilde{\mathbb{A}}\end{array}\right]=\left[\begin{array}{cc|cc} A&\Psi&B&\Sigma\\\overline{\Psi}&U&\overline{\Sigma}&M\\\hline\widetilde{B}&\widetilde{\Sigma}&\widetilde{A}&\widetilde{\Psi}\\\widetilde{\overline{\Sigma}}&\widetilde{M}&\widetilde{\overline{\Psi}}&\widetilde{U}\end{array}\right]\ ,   \label{X4x4}
\eea
the flatness condition \eq{Xflat} turns now into the analogues of \eq{Xflat1}-\eq{Xflat2}, but for $\bcFS$-valued forms,
\bea  d\mathbb{A}+\mathbb{A}\star \mathbb{A}+\mathbb{B}\star \widetilde{\mathbb{B}}= 0\ ,\qquad d\widetilde {\mathbb{A}}+\widetilde{\mathbb{A}}\star \widetilde{\mathbb{A}}+\widetilde{\mathbb{B}}\star {\mathbb{B}}= 0\ ,\label{Xbflat1}\\
d\mathbb{B}+\mathbb{A}\star \mathbb{B}-\mathbb{B}\star \widetilde{\mathbb{A}}= 0\ ,\qquad d\widetilde {\mathbb{B}}+\widetilde{\mathbb{A}}\star \widetilde{\mathbb{B}}-\widetilde{\mathbb{B}}\star {\mathbb{A}}= 0\ .\label{Xbflat2} \eea
The truncations of interest will only activate a few of the components of \eq{X4x4}, and the corresponding curvature equations \eq{Xbflat1}-\eq{Xbflat2} will accordingly take on different dynamical meanings. 

\subsection{HSG and CCHSG reductions}\label{sec:red}

Indeed, for a consistent truncation to correspond to HSG, as it is evident from \eq{Vas1}-\eq{Vas2}, $\mathbb{A}=\widetilde{\mathbb{A}}$, and the source term $\mathbb{B}\star \widetilde{\mathbb{B}}$ must be expanded purely on ${\rm End}(\cS)$. In fact, HSG is selected by requiring that only the ${\rm End}(\cS)$-sector is non-vanishing, and that in particular $\widetilde{\mathbb{B}}$ must reduce to $I_\Comp$ (more precisely, to $I_\Comp\star {\rm Id}_{\bcFS}$), which selects $SL(2,\Comp)$ as structure group (which can be projected out of $X$ via involutive automorphisms of the $\bcFS$  \cite{FSG1}). This way, \eq{Xbflat1}-\eq{Xbflat2} indeed reduce to \eq{Vas1}-\eq{Vas2},
\bea X = \left[\begin{array}{c|c}\mathbb{A}&\mathbb{B}\\\hline\widetilde{\mathbb{B}}&\widetilde{\mathbb{A}}\end{array}\right]=\left[\begin{array}{cc|cc} A&0 & B & 0 \\ 0 & 0& 0& 0\\ \hline \widetilde{B}&0 & A &0\\ 0 & 0 & 0 & 0 \end{array}\right]  \ \ \longrightarrow  \ \ \ \begin{array}{c} dA + A\star A + B\star I_{\mathbb{C}}  =  0 \ ,\\[5pt]  dB + A\star B - B\star A  =  0 \ . \end{array} \eea

On the other hand, in order to encode a 3D theory of conformal matter interacting with HS fields, 

\begin{enumerate}

\item the zero-form $\mathbb{B}$ must contain conformal matter fields; 

\item the two-form $\widetilde{\mathbb{B}}$ must break vacuum symmetries down to boundary Lorentz times dilations, $SL(2,\Real)\times O(1,1)_D$;

\item and $\mathbb{B}\star \widetilde{\mathbb{B}}$, candidate for containing conformal currents, must possess a non-trivial component admitting expansion over ${\rm End}(\cS)$ in order to couple to the gauge fields in $\mathbb{A}$. 

\end{enumerate}
A consistent truncation that ticks all boxes above and respects reality conditions on $X$ \cite{FSG1} is obtained via $\mathbb{A}=\widetilde{\mathbb{A}}$,  still both diagonal, while this time $\mathbb{B}$ as well as $\widetilde{\mathbb{B}}$ are off-diagonal, i.e., expanded over the mixed sectors, 
\bea X = \left[\begin{array}{c|c}\mathbb{A}&\mathbb{B}\\\hline\widetilde{\mathbb{B}}&\widetilde{\mathbb{A}}\end{array}\right]=\left[\begin{array}{cc|cc} W&0 & 0 & C \\ 0 & V & \ovC & 0\\ \hline 0 & C\star I_\Real  &  W & 0 \\ \ovC\star I_{\Real} & 0 & 0 & V \end{array}\right] 
\label{CCHSGdec}\eea
$W$ and $V$ are, respectively, a conformal HS gauge field connection and a colour one-form connection; the zero-form contains a mixed bimodule $C$ and its complex conjugate; while the two-form contains the product of $C$ with $I_\Real$, where the latter is a cohomologically non-trivial $SL(2,\Real)\times O(1,1)_D$-invariant two-form \cite{FSG1}.

As we are going to describe a CFT, it is useful to change coordinates on $\bY$ to the real combinations 
\be y^\pm_\a := \frac{e^{\pm i\pi/4}}{\sqrt{2}}(y_\alpha\mp i\bar{y}_{\dot \alpha})\ ,\ee
that act as raising/lowering operators with respect to the dilation operator $D$ \cite{F&L} (which can be embedded into $\mso(2,3)$ as any spacelike transvection \cite{FSG1,FSG2}),
\be [D,y^\pm_\alpha]_\star=\pm\frac{i}{2}\,y^\pm_\alpha \ , \qquad [y^+_\alpha, y^{-}_\beta]_\star  =  2i\epsilon_{\alpha\beta} \ .\ee
The $\mso(2,3)$ generators in the conformal slicing are given in terms of their star-product oscillator realization as 
\be T_{\alpha\beta}=\frac12\,y^+_\alpha y^+_\beta\ , \quad K_{\alpha\beta}=-\frac12\,y^-_\alpha y^-_\beta \ , \quad M_{\alpha\beta}=\frac12\,y^+_{(\alpha} y^-_{\beta)}\ ,\quad D= \frac14 \,y^{+\alpha} y^-_{\alpha} \ , \ee
where $T_{\a\b}$ generate (commuting) translations on ${\rm Mink}_3$, $K_{\alpha\beta}$  special conformal transformations and $M_{\a\b}$ the boundary Lorentz group. 
In a parallel fashion, we change coordinates on $\bZ$ to $z^\pm_\a$, defined analogously.

In these coordinates, the $SL(2,\Real)\times O(1,1)_D$-invariant two-form can a priori be taken as any of the two constructs
\be I^\pm_{\mathbb{R}}=\frac{i\pi}{2}\ dz^{\pm\alpha}dz^\pm_\alpha\ \delta^2_{\mathbb{C}}(z^\pm) \ , \ee
where $\delta^2_{\mathbb{C}}(z^\pm)$ is the analytic delta function defined in \cite{meta}. In fact, before assigning any vacuum expectation value to the one-forms, the $I^\xi_{\mathbb{R}}$  reductions are equivalent. This polarization symmetry is broken by the choice of the generator that the frame field gauges: gauging $y^{-\xi}_\alpha y^{-\xi}_\beta$, the $I^\xi_{\mathbb{R}}$ reduction leads to a perturbative description in terms of conformal matter fields; while gauging $y^{\xi}_\alpha y^{\xi}_\beta$ leads to a perturbative description in terms of dual, conformal matter fields \cite{BMV1}. In the following we shall thus restrict ourselves to reducing with $I^-_{\Real}$ and to encoding only conformal scalar fields in $C$. This also ensures that the star product $C\star \overline{C} \star I^-_{\mathbb{R}}$ is regular and that $I^-_{\mathbb{R}}$ is central in the DGA of the CCHSG system. 

Under \eq{CCHSGdec}, Eqs.  \eq{Xbflat1}-\eq{Xbflat2} decompose into
\bea  dW+W\star W + C\star \overline{C} \star I^-_{\mathbb{R}} &=& 0 \ ,\label{CCHSG1}\\
dC+W\star C-C\star V &= & 0\ ,\label{CCHSG2}\\
dV+V\star V+\overline{C}\star C \star I^-_{\mathbb{R}} &=& 0 \ . \label{CCHSG3} \eea 
As we recall below, $ C\star \overline{C} \star I^-_{\mathbb{R}}$ gives rise to a generating function for conformal currents. Thus, the system \eq{CCHSG1}-\eq{CCHSG3} shows immediately that consistency forces the addition of the colour gauge field $V$, sourced by the hermitian conjugate combination $\overline{C}\star C \star I^-_{\mathbb{R}}$, and can be thought of as a non-linear completion of the systems proposed in \cite{Misha} and \cite{Nilsson} (which did not make use of colour states nor internal gauge fields).

\subsection{Perturbative analysis of the CCHSG model}\label{sec:CCHSG}

The system \eq{CCHSG1}-\eq{CCHSG3} admits a one-parameter family of vacua singled out by $C^{(0)}=0=V^{(0)}$ and $W^{(0)}=\Omega \in \mathfrak{iso}(1,2)\oplus \mathfrak{so}(1,1)$, such that
\be d\Omega + \Omega\star \Omega = 0 \ ,\qquad \Omega = \Omega_x + \Omega_\rho \ , \ee
foliating $\bX_4$ with $d_X=d_x+d_\r$ and
\be \Omega_x = -i\left(\frac{1}{\rho}e^{\alpha\beta}T_{\alpha\beta}+ \frac12 \omega^{\alpha\beta}M_{\alpha\beta}\right)\ , \quad \Omega_\rho = -i\frac{d\rho}{\rho}D \ , \qquad \rho \in \mathbb{R}^+ \ ,\ee
corresponding, from the bulk point of view, to the standard Poincar\'e patch metric $ds^2=(dx^2+d\r^2)/\r^2$. Expanding in the zero-form around such vacuum yields
\bea
    &dC^{(1)} +\Omega\star C^{(1)}=0\ ,\qquad d\overline C^{(1)} -\overline C^{(1)}\star \Omega =0\ ,&\label{C1xR}\\
    &d W^{(2)}+  {\Omega}\star W^{(2)}+W^{(2)}\star \Omega + C^{(1)}\star \overline{C}^{(1)}\star I^-_{\Real}=0\ ,&\label{W2xR}\\
    &d V^{(2)}+ \overline C^{(1)}\star {C}^{(1)}\star I^-_{\Real}=0\ .&\label{V2xR}
\eea

As $I^-_\Real$ does not contain any Klein element with non-trivial action on $Y$, we can restrict to a unitary internal 3D model by taking $\cV$ to be a $U(N)$-module. Moreover, the realization of the background connection $\O_x$ selects, as Poincar\'e-invariant vacuum, the conformal singleton highest-weight state $\ket{(-i/2)}$ (see \cite{FSG1,FSG2} for details on this representation) and implies that a standard 3D conformal scalar be expanded over states $(y^-)^n_{\a_1...\a_n}\star\ket{(-i/2)}$. 
Thus, a conformal scalar solution of \eq{C1xR} is encoded in the linearized intertwining zero-form master fields
\be C^{(1)}= \sqrt{\r}\,c^{(1) {I}}\star \ket{ (-i/2)}\bra{e^+_{I}}\ , \qquad \overline C^{(1)}=\sqrt{\r} \,\ket{ e^{+, { I}}}\bra{ (i/2)}\star\, \bar c^{(1)}_{I}\ , \label{expC}\ee
where\footnote{We use $m,n$ for 3D Lorentz indices, with the usual map to spinor indices $\a,\b$ implemented via 3D van der Waerden symbols $(\gamma_m)^{\a\b}$, $x^{\a\b}=x^{\b\a}=x^m(\gamma_m)^{\a\b}$ (see Appendix A.2 in \cite{FSG1} for the details). We also use the shorthand notation $T_{m(n)}$ to denote a tensor with $n$ totally symmetrized indices $T_{m_1...m_n} =
T_{(m_1...m_n)}$, both for vector and spinor indices. Repeated non-contracted indices are also to be understood as totally symmetrized, $S_{\a(2)}T_{\a(2)}=S_{\a\a}T_{\a\a} :=
S_{(\a_1 \a_2} T_{\a_3 \a_4)}$.}
\be  c^{(1){I}}=\sum_{n=0}^\infty \frac{\rho^n}{n!}\,\phi^{{I},m(n)} K_{m_1}...K_{m_n}= \sum_{n=0}^\infty \frac{1}{ (2n)!}\,\phi^{ {I},\alpha(2n)}(x)\,q^-_{\alpha_1}\ldots q^-_{\alpha_{2n}} \ ,  \label{cI} \ee
with $ q^-_\alpha:=\sqrt{\rho}\,y^-_\alpha$, and analogously for the complex conjugate $\bar c^{(1)}_{I}$. Eq. \eq{cI} is a generating function for an unfolded 3D complex massless scalar field. Indeed, \eq{C1xR} imposes the unfolded chain of equations
\be \nabla\phi^ {I}_{m(n)} - i \, e^p\,\phi^ {I}_{m(n)p} = 0 \ , 
\ee
where $\nabla\equiv \nabla_x$, imposing the 3D Klein-Gordon equation $\nabla^2 \phi^{I} = 0$ on the physical scalar $\phi^I$ and solving all higher-rank tensors in \eq{cI} in terms of its derivatives. 

We now turn to looking at the gauge-field source term $C^{(1)}\star \overline{C}^{(1)}\star I^-_{\Real}$, expected to generate conformal currents. First, it is crucial that the star products involved actually give rise to a regular element. Indeed, using the orthonormality of colour states, one can show that
\bea & C^{(1)}\star\overline{C}^{(1)}\star I^-_{\mathbb{R}} = \rho\, c^{(1) { I}}\star \underbrace{|{ (-i/2)}\rangle \langle{(i/2)}|}_{=\,4\pi \delta^{2}_{\mathbb{C}}(y^+)} \star I^-_{\mathbb{R}}\star  \bar{c}^{(1)}_I &\nonumber\\ 
& =   \frac{i}{2}\, dz^{-\a}\wedge dz^-_\a\, \rho\, c^{(1) I}\star \exp(iy^+ z^-) \star \bar{c}_ {I}^{(1)} =: \frac{i}{2}\,\rho\, dz^{-\a}\wedge dz^-_\a\,{\cal J}\ , & \eea 
where ${\cal J}={\cal J}(x,z^-,y^+,y^-)$ is a regular function of the oscillators, and such that its projection onto $z^-=0$ contains 3D conformal currents of all spins \cite{Anselmi,Konstein:2000bi,Giombirev},
\be \left.{\cal J}\right|_{z^-=0}= \sum_{s=0}^\infty \frac{\rho^s}{s!}\,{\cal J}_{s}^{m(s)}(x) K_{m_1}...K_{m_s}  = \sum_{s=0}^\infty \frac{\rho^s}{(2s)!}\,{\cal J}_{s}^{\alpha(2s)}(x) y^-_{\alpha_1}...y^-_{\alpha_{2s}} \ , \ee
with 
\be {\cal J}_{s,m(s)} \ = \ 
i^s\sum_{k=0}^s {2s \choose 2k}(-1)^k\,\nabla_{\{m(k)}\phi^ { I}\, \nabla_{m(s-k)\}}\bar\phi_ { I} \ , \ee
where curly brackets enclosing indices denote symmetric traceless projection. Thus, integrating with respect to $z^-$ with the simplest contracting homotopy \cite{FSG2} and projecting onto $z^-=0$, the physical content of \eq{W2xR} reduces to the two conditions 
\bea  D_x w^{(2)}_x  & = & i\,e^{\alpha\gamma}e_\gamma{}^\beta\,\partial_{q^{-\alpha}} \partial_{q^{-\beta}}  \left.{\cal J}\right|_{z^-=0} \ ,\\
D_{\rho} w^{(2)}_x & = & 0\ ,\eea
the first one determining the coupling of conformal HS gauge fields in $w^{(2)}_x:=W_x^{(2)}|_{z^-=0}$ with the primary currents in $\left.{\cal J}\right|_{z^-=0}$, and the second one determining the scale dependence of the gauge fields.

The hybrid nature \eq{expC} of $C^{(1)}$ is key to encoding matter currents in a star-factorized construct of the 3D master fields. However, it also generates a backreaction on the colour gauge field, since the source term in \eq{V2xR} admits expansion over the colour sector and is non-trivial,
\bea \overline C^{(1)}\star {C}^{(1)}\star I^-_{\Real} &=& \ket{e^{+,I}}\bra{(i/2)}\star \bar c_I^{(1)}\star c^{(1)J} \star \ket{(-i/2)}\bra{e^+_J}\star I^-_{\Real} \nonumber\\ & =:& \ket{e^{+,I}} \bra{e^+_J}U_I{}^J\star I^-_{\Real}\ .\label{Vsource}\eea
In order to compute the colour matrix elements $U_I{}^J$ it is convenient to expand the scalar field $c^{(1)I}$ on a basis of states which diagonalises $y^+$, and thus the 3D energy, and with well-defined inner product, like the momentum-eigenstate basis $e^{\tfrac{i}2\l y^-}  \star \ket{(-i/2)}=\ket{+;\e;\l}$ \cite{FSG1,FSG2}. This way, we can separate positive-frequency ($c^{(1)I}_>$, expanded over $\l_\a\in\Real^2$) and negative-frequency ($c^{(1)I}_<$, expanded over $\l_\a\in i\Real^2$) components of the 3D scalar field, and expand them in terms of real momenta $\ell_\a$ as 
\be  c^{(1)I}_> := \int_{\Real^2} \frac{d^2\ell}{4\pi}\,\widetilde{\phi}^I_>(\ell,x)\,e^{\tfrac{i}2\ell y^-} \ , \qquad c^{(1)I}_< := \int_{\Real^2}  \frac{d^2\ell}{4\pi}\,\widetilde{\phi}^I_<(\ell,x)\,e^{-\tfrac{1}{2}\ell y^-} \ , \label{cfreq} \ee
with transforms $\widetilde{\phi}^I_{\lessgtr}(\ell,x)$ being even functions of $\ell_\a$. The star products in \eq{Vsource} give rise to a regular source \cite{FSG2}, with \emph{non-local} colour current
\be U_I{}^J = 4\int d^2\ell \,\left[\widetilde{\bar \phi}_{I>}\,\widetilde{\phi}^J_>-\widetilde{\bar \phi}_{I<}\,\widetilde{\phi}^J_<\right] \ .\ee

Thus, the CCHSG system candidate dual to HSG naturally incorporates a minimally coupled $U(N)$ gauge field $V$ --- in this sense resembling an ordinary colour gauge field. The fractional-spin algebra origin of the system implies that the latter cannot be truncated away consistently, except under technical assumptions which seem unnatural (see Section 4.7 in \cite{FSG2} for a discussion). The source term for $V$ however is a non-local construct of the scalar field: this distinguishes $V$ from an ordinary 3D colour Chern-Simons gauge field \cite{vecCS}, sourced by a local current. We stress also that the colour generators $\ket{e^{+,I}} \bra{e^+_J}$ are not external but admit realization on $\bY$, and therefore the interaction of colour and conformal sectors is controlled by a single star-product algebra. In this sense, we can think of the colour degrees of freedom in the CCHSG system as dynamically realized Chan-Paton factors.

\subsection{Fractional-spin gravity}

Descending from the full parent system \eq{Xbflat1}-\eq{Xbflat2} to Vasiliev's HSG, one passes via an intermediate 4D system with richer field content, obtained by the consistent reduction 
\bea \widetilde{\mathbb{A}}={}&\mathbb{A}=\left[\begin{array}{cc} A & \Psi\\ \overline \Psi & V\end{array}\right] \ , \quad \mathbb{B} = \left[\begin{array}{cc} B &\Xi\\ \overline \Xi& N\end{array}\right]\ ,\quad \widetilde{\mathbb{B}}= I_\Comp\star\left[\begin{array}{cc}
  {\rm Id}_{{\cal S}}   &0  \\
   0  & {\rm Id}_{{\cal V}}
\end{array}\right] \ ,\eea
in which both the zero- and one-forms consist of diagonal components, encoding integer-spin and colour-adjoint fields, \emph{and} off-diagonal components, encoding coloured singletons (as seen in Section \ref{sec:CCHSG}).

As a result, the parent flatness condition now reduces to the integrable field equations
\be d\mathbb{A}+\mathbb{A}\star \mathbb{A}+\mathbb{B}\star I_{\Comp}= 0\ ,\qquad d\mathbb{B}+\mathbb{A}\star \mathbb{B}-\mathbb{B}\star {\mathbb{A}}= 0 \ .\ee
Their linearization around the $AdS_4$ background (see Section 3.5 in \cite{FSG2}) describes the coupling of fields $(\Psi,\Xi)$, with fractional $SL(2,\Comp)$ spin, to Vasiliev's pure HSG master fields $(A,B)$ and to the internal colour sector $(V,N)$. We refer to this system as \emph{fractional-spin gravity} (FSG). The crucial technical point that enables the coupling of singletons to bulk HSG and colour integer-spin fields is the simultaneous presence, due to the $\bcFS$-algebra expansion, of one-sided as well as two-sided modules of the background isometry algebra in both the one-form and the zero-form of the system. 

The inclusion of singletons into the ``bulk branch'' of the parent system requires, however, a non-unitary extension of the colour module with maximally split signature, in which $\kappa_y$, included in $I_\Comp$, acts faithfully (as explained in Section \ref{sec:FS}): therefore, in this case ${\rm End}(\cV) = U(N,N)$. Considering the bulk singletons as noncommutative analogues of D-branes, this extension is analogous to the need for D-branes and anti-D-branes in supergravity --- an analogy which we would like to make more precise. Moreover, the treatment of singletons as exotic bulk matter fields that cannot show up as localisable particles while backreacting to the bulk gauge fields, including gravity, may provide a viable extended QFT framework including dark matter along the lines of the early proposal \cite{Flato:1980zk,Flato:1986uh} and its later refinement in \cite{Vasiliev:2025ocb}. Finally, it would be interesting to investigate the role of FSG within our holography framework.

\begin{acknowledgement}

It is a pleasure to thank the Organizers for arranging a very stimulating workshop. The results reported on in this paper have been obtained in collaboration with F. Diaz. We are grateful to M. A. Vasiliev for useful discussions. C.I. also wishes to thank S. Afentoulidis-Almpanis, R. Grady, J. Lang, V. Letsios, D. Ponomarev, E. Skvortsov, R. Van Dongen for helpful exchanges. P.S. also acknowledges useful conversations with M. Valenzuela and support received from UNAP -- VRII Consolida grant ``Higher-spin inspired IR modifications of 3d gravity''; ANID grant Regular N. 1250672, ``Reinventing Quantum Field Theory: Higher-Spin-Inspired Modifications of
Gravity, Gauge Theory and Holography''; and the MATH-AMSUD project SGP 24-MATH-12, ``Symmetries in Geometry and Physics''. co-funded by ANID and Minist\`ere de l'Europe et des Affaires \'Etrang\`eres.

\end{acknowledgement}

\end{document}